\documentclass[twocolumn,aps,showpacs,nofootinbib,prd]{revtex4-1}
\usepackage{amssymb}
\usepackage{amsmath}
\usepackage{float}
\usepackage[dvips]{color}
\usepackage{graphicx}
\usepackage{color}
\usepackage{soul}
\DeclareMathOperator{\tr}{\mbox{Tr}}

\newcommand{\be}{\begin{equation}}
\newcommand{\ee}{\end{equation}}
\newcommand{\bea}{\begin{eqnarray}}
\newcommand{\eea}{\end{eqnarray}}
\newcommand{\beas}{\begin{eqnarray*}}
\newcommand{\eeas}{\end{eqnarray*}}

\newcommand{\slsh}[1]{{\not \! #1}}

\begin{document}
\title{Thermo-magnetic properties of the strong coupling in the local Nambu--Jona-Lasinio model}
\author{Alejandro Ayala$^{1,2}$, C. A. Dominguez$^2$, L. A. Hern\'andez$^2$, M. Loewe$^{2,3,4}$, Alfredo Raya$^5$, J. C. Rojas$^6$, C. Villavicencio$^7$}
\affiliation{$^1$Instituto de Ciencias
  Nucleares, Universidad Nacional Aut\'onoma de M\'exico, Apartado
  Postal 70-543, M\'exico Distrito Federal 04510,
  Mexico.\\
  $^2$Centre for Theoretical and Mathematical Physics, and Department of Physics,
  University of Cape Town, Rondebosch 7700, South Africa.\\
  $^3$Instituto de F\1sica, Pontificia Universidad Cat\'olica de Chile,
  Casilla 306, Santiago 22, Chile.\\
  $^4$Centro Cient\1fico-Tecnol\'ogico de Valpara\1so, Casilla 110-V, Valpara\1so, Chile.\\
  $^5$Instituto de F\1sica y Matem\'aticas, Universidad Michoacana de San Nicol\'as de Hidalgo, 
  Edificio C-3, Ciudad Universitaria, Morelia, Michoac\'an 58040, Mexico.\\
  $^6$Departamento de F\1sica, Universidad Cat\'olica del Norte, Casilla 1280, Antofagasta, Chile\\
  $^7$Departamento de Ciencias B\'asicas, Facultad de Cienicas, Universidad del B\'io-B\'io,
  Casilla 447, Chill\'an, Chile.}

\begin{abstract}

We study the thermo-magnetic properties of the strong coupling constant $G$ and quark mass $M$ entering the Nambu-Jona-Lasinio model. For this purpose, we compute the quark condensate and compare it to lattice QCD (LQCD) results to extract the behavior of $G$ and $M$ as functions of the magnetic field strength and temperature. We find that at zero temperature, where the LQCD condensate is found to monotonically increase with the field strength, $M$ also increases whereas $G$ remains approximately constant. However, for temperatures above the chiral/deconfinement  phase transitions, where the LQCD condensate is found to monotonically decrease with increasing field, $M$ and $G$ also decrease monotonically. For finite temperatures, below the transition temperature, we find that both $G$ and $M$ initially grow and then decrease with increasing field strength. To study possible consequences of the extracted temperature and magnetic field dependence of $G$ and $M$, we compute the pressure and compare to LQCD results, finding an excellent qualitative agreement. In particular, we show that the transverse pressure, as a function of the field strength, is always negative for temperatures below the transition temperature whereas it starts off being positive and then becomes negative for temperatures above the transition temperature, also in agreement with LQCD results. We also show that for the longitudinal pressure to agree with LQCD calculations, the system should be described as a diamagnet. We argue that the turnover of $M$ and $G$ as functions of temperature and field strength is a key element that drives the behavior of the quark condensate going across the transition temperature and provides clues for a better understanding of the inverse magnetic catalysis phenomenon.
 
\end{abstract}


\maketitle

\section{Introduction}\label{I}

The study of the properties of strongly interacting matter at high temperature and/or density, under the influence of magnetic fields, has become a research subject of 
growing interest over the last years. Examples of physical systems where such properties are relevant include heavy-ion collisions and compact stellar objects. One of 
the intriguing aspects of these properties, not yet well understood (at least in a consensual way), are the possible causes of the so called {\it inverse magnetic 
catalysis} (IMC) phenomenon found by lattice QCD (LQCD) calculations~\cite{bali1,bali2,bali3}. Recall that IMC is characterized by a decreasing critical temperature 
($T_c$) for the chiral/deconfinement phase transition and a decreasing quark condensate above $T_c$, with increasing field strength. 

Different approaches have been explored in order to either find or include IMC in QCD~\cite{attention}, some of these provide an explanation~\cite{explains}, and almost
all suggest the need to include extra ingredients in terms of magnetic-induced modifications of QCD properties. In particular, the modification of the QCD coupling due 
to magnetic screening at low temperature and antiscreening at high temperature has been shown to be a plausible mechanism to explain 
IMC~\cite{decreasing1,decreasing,coupling,decreasing2}. This also seems to be the reason why effective models without such modifications do not describe neither the 
behavior of the critical temperature nor the properties of the quark condensate at high 
temperature~\cite{Boomsma01,Loewe1,Agasian,Fraga,Fraga2,Andersen,Blaschke01,Scoccola01}. For recent reviews see~\cite{review1,review2}. 

Deducing the detailed screening/antiscreening properties of the strong coupling as a function of the field strength is not a simple task since these properties belong 
for the largest portion of the parameter space to the non-perturbative domain. Nevertheless, it should be possible to extract general features of this coupling by 
resorting to combining information from effective models and LQCD. 

The Nambu--Jona-Lasinio (NJL) is one of such models. It has been extensively used to explore the chiral transition~\cite{Buballa01,Klevansky01}. In particular, the NJL model can be used to formulate a simplified version of the QCD gap equation by means of the Schwinger-Dyson technique where the dynamically generated mass $M$ is constant (momentum independent) and the interaction is given by a four-fermion contact term whose strength is controlled by a coupling $G$.  A pertinent question is whether it is possible to extract information on the behavior of the coupling $G$ as a function of the magnetic field strength and the temperature by combining the NJL model with LQCD data for the quark  condensate in the presence of a magnetic field, and whether this information can be used to get clues on the microscopical origins of IMC. 

In this work we take this approach. We use the NJL model to extract the behavior of $G$ and $M$ as functions of the magnetic field for different temperatures using LQCD data for the quark condensate~\cite{bali2}. Notice that if there is a link between the fading-out of the condensate, as a function of $eB$, above the critical temperature for the chiral/deconfinement transition, and the thermo-magnetic dependent coupling, then the latter should also decrease with the field intensity. This behavior would signal that a decreasing coupling contributes to a less intense bound between quark-antiquark pairs above the critical temperature, as the field intensity grows. In this work we show that this is the case. 

The work is organized as follows: In Sec~\ref{II} we set up the framework writing the expression for the gap equation and the quark propagator obtained from the NJL  model in the constant (momentum independent) mass approximation. These equations, together with the expression for the quark condensate given in terms of the  quark propagator, provide the set of equations that allow finding the behavior of $M$ and $G$ as functions of the field strength, for different temperatures. In  Sec.~\ref{III} we include the effects of the magnetic field by means of Schwinger's proper time method. Since the NJL model is not renormalizable, in order to find the  behavior of $M$ and $G$ as functions of the magnetic field, we first  separate the vacuum contribution from the thermo-magnetic one. This procedure has been shown to give reliable results for matter and magnetic field induced properties in the NJL model~\cite{Norberto}. In Sec.~\ref{IV} we compute the thermo-magnetic dependence of $M$ using as input the LQCD behavior of the average light quark condensate as a  function of the field strength. We also find the value of $G$ from the gap equation. Since the light quarks have different charges, we find the values for $G$ and $M$ by averaging over the light-quark flavors. We also compute the thermo-magnetic contribution to the pressure and show that above $T_c$ the transverse pressure starts off being positive for small field strengths. Below $T_c$ this pressure starts from zero and then becomes negative as the field strength increases, in agreement with LQCD calculations.  We argue that this result goes in line with the idea that above (below) $T_c$ quarks are brought together (pushed apart) and this makes the coupling become weaker  (stronger) due to asymptotic freedom. We also explore the sensitivity of the results for the pressure to variations of the vacuum parameters and find that there is no significant dependence. We finally summarize and conclude in Sec.~\ref{concl}.

\section{Gap equation and quark condensate in the NJL model}\label{II}

The NJL model is defined by means of the Lagrangian density
\begin{equation}
\mathcal{L}=\bar{\psi}(i\slsh{\partial}-m)\psi+G\left(\left(\bar{\psi}\psi\right)^2+\left(\bar{\psi}i\gamma^5\boldsymbol{\tau}\psi\right)^2\right)\label{lagrangiannjl},
\end{equation}
where $\boldsymbol{\tau}$ are the Pauli matrices in isospin space, and $\psi$ is a quark field. 

On general grounds, in the mean field approximation, and after a bosonization process, the Lagrangian can be rewritten as a vacuum term plus a free fermion Lagrangian with a dressed mass, namely
\begin{equation}
\mathcal{L}_{MF}=-\frac{\sigma^2}{4G}+\bar{\psi}(i\slsh{\partial}-M)\psi\;,
\end{equation}
where $\sigma = 4G\langle\bar\psi\psi\rangle$ and $M=m+\sigma$. Here we do not consider pion condensation effects, so the only contribution comes from the sigma meson in the bosonization procedure.
The value of the mean field is determined through the {\em gap equation}, obtained by minimizing the effective potential with respect to the mean field \cite{Buballa01,Klevansky01}
\begin{equation}
M-m=4G\int\frac{d^4p}{(2\pi)^4}\tr[iS(p)]\label{gap1},
\end{equation}
with the trace referring to color and Lorentz indices. We notice that the quark condensate $\langle \bar{\psi} \psi\rangle$  is given by 
\begin{eqnarray} 
\langle \bar{\psi}\psi\rangle = -\int\frac{d^4p}{(2\pi)^4}\tr[iS(p)].
\label{qbarq}
\end{eqnarray}
In the absence of thermo-magnetic effects, the propagator is given by
 \bea
   S(p)=\frac{\slsh{p}\, + M_0}{p^2-M_0^2 +i\epsilon}.
\eea
We now proceed to include magnetic field and temperature effects in the model.

\section{thermo-magnetic effects}\label{III}

Eqs.~(\ref{gap1}) and~(\ref{qbarq}) represent the two independent equations providing information on the thermo-magnetic behavior of the coupling $G$ and the 
dynamically generated mass $M$, after using  LQCD results for the quark condensates~\cite{bali2}. 

To account for the  magnetic field, we emphasize that the above described bosonization does not affect the form of the gap equation~(\ref{gap1}) nor the 
condensate~(\ref{qbarq}), and the effect of the magnetic field is reflected in the dressing of the quark propagator. For this we resort to Schwinger proper time representation of the  two-point function
 \begin{eqnarray}
&& i S(p) = \int _{0}^{\infty}\frac{ds}{\cos(q_fBs)}e^{is(p_{\parallel}^{2} - p_{\perp}^{2}\frac{\tan (q_fBs)}{q_fBs}- M^{2} +i\epsilon)}\nonumber\\
&&\biggl[\left(\cos (q_fBs) + \gamma_1 \gamma_2 \sin (q_fBs)\right)
   (M+\slsh{p_{\|}}) - \frac{\slsh{p_\bot}}{\cos(q_fBs)} \biggr],\nonumber\\
   \label{tracewithSchwinger}
 \end{eqnarray}
where $q_f$ is the absolute value of the quark charge (i.e. $q_u = 2|e|/3$ and $q_d = |e|/3$), and we have chosen the homogeneous magnetic field to point in the 
$\hat{z}$ direction, namely $\boldsymbol{B}=B\hat{z}$. This configuration can be obtained from an external vector potential which we choose in the so called {\it symmetric gauge}
\begin{equation}
A^{\mu}= \frac{B}{2}(0,-y,x,0).
\end{equation}
We have also defined
\bea
p_\perp^\mu&\equiv&(0,p_1,p_2,0),\nonumber\\
p_\parallel^\mu&\equiv&(p_0,0,0,p_3),\nonumber\\
p_\perp^2&\equiv& p_1^2+p_2^2,\nonumber\\
p_\parallel^2 &\equiv& p_0^2-p_3^2.
\label{pps}
\eea
Notice that since the magnetic field breaks Lorentz invariance, the propagator involves a non-local, albeit path independent phase. However, by taking the trace over 
a closed one-loop diagram, as is required for the calculation of the condensate, this phase does not contribute and thus we ignore it in the sequel.

Using Eq.~(\ref{tracewithSchwinger}) to take the trace in Eq.~(\ref{qbarq}), we obtain
\bea
   \langle \bar{\psi} \psi\rangle&=&-4N_cM\frac{1}{2}\sum_{f}\int\frac{d^4p}{(2\pi )^4}
   \nonumber\\
   &\times&\int _{0}^{\infty}dse^{is(p_{\parallel}^{2} - p_{\perp}^{2}\frac{\tan (q_fBs)}{q_fBs}- M^{2} +i\epsilon)},
\label{takingtrace}
\eea
where in order to account for the different quark charges $q_f$ we have averaged over quark-flavors.

The integration over the transverse momentum components can be carried out, leading to
\begin{equation}
\int \frac{d^{2}p_{\perp}}{(2\pi )^{2}}e^{-i \frac{\tan (q_fBs)}{q_fB}p_{\perp}^{2}} = \frac{q_fB}{4\pi i}\frac{1}{\tan (q_fBs)}.
\label{intoverperp}
\end{equation}

In order to introduce a finite temperature, within the Matsubara formalism, we transform the integrals to Euclidean space by means of
\begin{equation}
\int \frac{d^{2}p_{\parallel}}{(2\pi)^{2}} \to iT \sum_{n= -\infty}^{+\infty}\int \frac{ dp_{3}}{(2\pi)},
\label{tang}
\end{equation}
where the integral over the zeroth component of the fermion momentum has been discretized. We also 
perform the change of variable $s = -i\tau $. Therefore, the expression for the quark condensate in Eq.~(\ref{takingtrace}) becomes
\begin{eqnarray}
\langle \bar{\psi}\psi\rangle &=& -N_{c}M\frac{1}{2}\sum_{f}\frac{q_fB}{\pi}T\sum_{n=-\infty}^{+\infty}\int_{-\infty}^{+\infty}\frac{dp_{3}}{(2\pi)}
\nonumber\\
&\times&\int_{\tau_{0}}^{+\infty}\frac{d\tau}{\tanh(q_fB\tau)}
e^{-\tau(\tilde{\omega_{n}^2} + \omega ^{2}) },
\label{inter}
\end{eqnarray}
where we have introduced the fermion Matsubara frequencies $\tilde{\omega}_{n} = (2n+1)\pi T$ and $\omega^2=p_3^2 + M^2$.

Since the NJL model is non-renormalizable the integral above needs to be regularized. This can be done in different ways,  the simplest being the introduction of an ultraviolet cut-off. This is tantamount to introducing a regulator as the lower limit cutoff in the proper-time representation. The parameter $\tau_{0}$ represents such regulator.

The sum in Eq.~(\ref{inter}) can be expressed in terms of Jacobi's $\vartheta _{3}(z,x)$ function, defined as
\begin{equation}
\vartheta _{3} (z,x) = \sum_{n= -\infty}^{+\infty}\exp(i\pi x n^{2} +2i\pi z n),
\end{equation}
whereby
\begin{eqnarray}
\!\!\!\!\!\sum_{n=-\infty}^{+\infty} e^{-\tau(2n+1)^{2}\pi^{2}T^{2}} = e^{-2\pi ^{2}\tau T^{2}}
\vartheta_{3}(2\pi i\tau T^{2}, 4\pi i \tau T^2). \nonumber \\ 
\label{theta3}
\end{eqnarray} 

For our purposes it is useful to invoke the inversion formula 
\begin{equation}
\vartheta _{3} (z,x) = \sqrt{\frac{i}{x}}\exp\left(\frac{z^{2}\pi}{ix}\right)\vartheta _{3}\left(\frac{z}{x},-\frac{1}{x}\right),
\end{equation}
which leads to
\begin{eqnarray}
\langle \bar{\psi}\psi \rangle &=& -\frac{N_{c}M}{4\pi ^{2}}\frac{1}{2}\sum_{f} q_fB \int _{\tau_{0}} ^{\infty}\frac{d\tau}{\tau \tanh(q_fB\tau)}e^{-\tau M^{2}}\nonumber\\
&\times& \vartheta _{3}\left(\frac{1}{2}, \frac{i}{4\tau\pi T^{2}}\right).
\label{condthermmagn}
\end{eqnarray}
Using 
\begin{eqnarray}
\!\!\!\!\!\vartheta _{3}\left(\frac{1}{2}, \frac{i}{4\pi \tau T^{2}}\right) &=& 
\sum _{n = -\infty}^{+ \infty} (-1) ^n \exp \left(-\frac{n^{2}}{4\tau T^{2}}\right)\nonumber\\
&=&1+2\sum _{n = 1}^{+ \infty} (-1)^n \exp \left(-\frac{n^{2}}{4\tau T^{2}}\right),
\end{eqnarray}
we notice that the $T=0$ term corresponds to $n=0$ in the above expression. Therefore, the vacuum contribution is obtained from the $n=0$ term in the limit where $q_fB\to 0$. Adding and subtracting 1 in the integrand we get
\begin{eqnarray}
\langle \bar{\psi}\psi \rangle &=& -\frac{N_{c} M}{4\pi ^{2}}
\Big\{\int _{\tau _{0}}^\infty \frac{d \tau}{\tau ^{2}} e^{-\tau M^{2}}\nonumber\\
&+&
\frac{1}{2}\sum_{f} q_fB \int _{\tau_{0}} ^{\infty}\frac{d\tau}{\tau^2}e^{-\tau M^{2}}\left[\frac{q_fB\tau}{\tanh(q_fB\tau)}-1\right]\nonumber\\
&+&
\sum_{f} q_fB \sum_{n=1}^{\infty}(-1)^n 
\int _{\tau_{0}}^{\infty}d\tau \frac{e^{-\tau M^{2}}e^{-\frac{n^{2}}{4\tau T^{2}}}}{\tau \tanh(q_fB\tau)}\Big\},
\label{condthermmagnsep}
\end{eqnarray}
where we can identify the vacuum condensate as given by the expression
\begin{equation} 
  \langle \bar{\psi}\psi\rangle_{0} = -\frac{N_{c} M_0}{4\pi ^{2}}\int _{\tau _{0}}^\infty \frac{d \tau}{\tau ^{2}} e^{-\tau M_0^{2}},
  \label{conpropdvac}
\end{equation}
whereas the thermo-magnetic contribution is given by
\begin{eqnarray}
   \langle \bar{\psi}\psi\rangle_{B,T} &=& -\frac{N_{c} M}{4\pi ^{2}}\frac{1}{2}\sum_{f} q_fB 
   \Big\{ \nonumber\\
&&
   \int _{0}^{\infty}\frac{d\tau}{\tau^2}e^{-\tau M^{2}}\left[\frac{q_fB\tau}
   {\tanh(q_fB\tau)}-1\right]
   \nonumber\\
&+&
   2\sum_{n=1}^{\infty}(-1)^n 
   \int _{0}^{\infty}d\tau \frac{e^{-\tau M^{2}}e^{-\frac{n^{2}}{4\tau T^{2}}}}{\tau \tanh(q_fB\tau)}\Big\}. 
\label{condpropthermmag}
\end{eqnarray}
The quantity $M\equiv M(B,T)$ in Eq.~(\ref{condpropthermmag}) is such that when $B,T\to 0$, $M\to M_0$. It turns out that the integrals in Eq.~(\ref{condpropthermmag}) are finite as the lower limit of integration goes to zero. This means that the thermo-magnetic effects are independent of the regulator $\tau_0$ and we have consequently set this lower limit to zero in Eq.~(\ref{condpropthermmag}). Also, the first integral in this equation can be computed analytically, with the result
\begin{eqnarray}
   && \int _{0}^{\infty}\frac{d\tau}{\tau^2}e^{-\tau M^{2}}\left[\frac{q_fB\tau}
   {\tanh(q_fB\tau)}-1\right]
   = 2\left\{\frac{M^2}{2q_fB}  \right. \nonumber\\
   &&  - \frac{1}{2}\ln\left(\frac{M^2}{2q_fB}\right) 
   - \left(\frac{M^2}{2q_fB}\right)\ln\left(\frac{M^2}{2q_fB}\right)\nonumber\\
   &&  + \left. \ln\Gamma\left[1 + \frac{M^2}{2q_fB}\right]\right\}\!.
\label{analytical}
\end{eqnarray}

Equation~(\ref{conpropdvac}) can be used to fix the vacuum values of the quark condensate and the dynamically generated mass $M_0$ from a choice of $\tau_0$. Two sets of consistent choices, that reproduce the physical values of the pion mass and of the pion decay constant~\cite{Norberto} are shown in Table 1. Shown also are the corresponding vacuum values for the coupling constant $G_0$ and the current quark mass $m$. 

\begin{table}[t!]
\centering
\begin{tabular}{cccccc}
\hline
  $\tau_0$ & $-\langle\bar{\psi}\psi\rangle_0^{1/3}$ & $M_0$ & $G_0$ & $m$ & $T_c^{NJL}$\\
  (GeV)$^{-2}$ & (GeV) &  (GeV) & (GeV)$^{-2}$ &  (GeV) & (GeV) \\ \hline
 1.27 & 0.220 & 0.224 & 5.08 & 0.00758 & 0.267 \\ \hline
 0.74 & 0.260 & 0.192 & 2.66 & 0.00465  & 0.228
\end{tabular}
\caption{Two sets of values for the vacuum regulator $\tau_0$, condensate $\langle\bar{\psi}\psi\rangle_0$ and dynamically generated mass $M_0$ stemming from requiring that the pion mass and the pion decay constant computed in the NJL model attain their physical values. Shown also are the corresponding vacuum values for the coupling constant $G_0$, current quark mass $m$ and the critical temperature for $eB=0$.}
\label{table1}
\end{table}

Before proceeding to use Eqs.~(\ref{condpropthermmag}) and~(\ref{analytical}) to find the thermo-magnetic behavior of the dynamically generated mass $M$ and the coupling $G$ and the consequences for the pressure, it is important to notice that the critical temperature given by the model depends on the choice of $M_0$ and does not coincide with the corresponding value reported by lattice. It is therefore necessary to scale the values of the model temperatures to make them correspond to the physical values. The simplest choice is a linear scaling such that
\begin{eqnarray}
  T^{NJL}=\left(\frac{T_c^{NJL}}{T_c}\right)T,
\label{scaling}
\end{eqnarray}
where $T$ represents the physical value of the temperature and $T_c$ and $T_c^{NJL}$ are the physical and model critical temperatures, respectively. 

\begin{widetext}
\begin{figure*}[t]
\begin{center}
\includegraphics[scale=0.42]{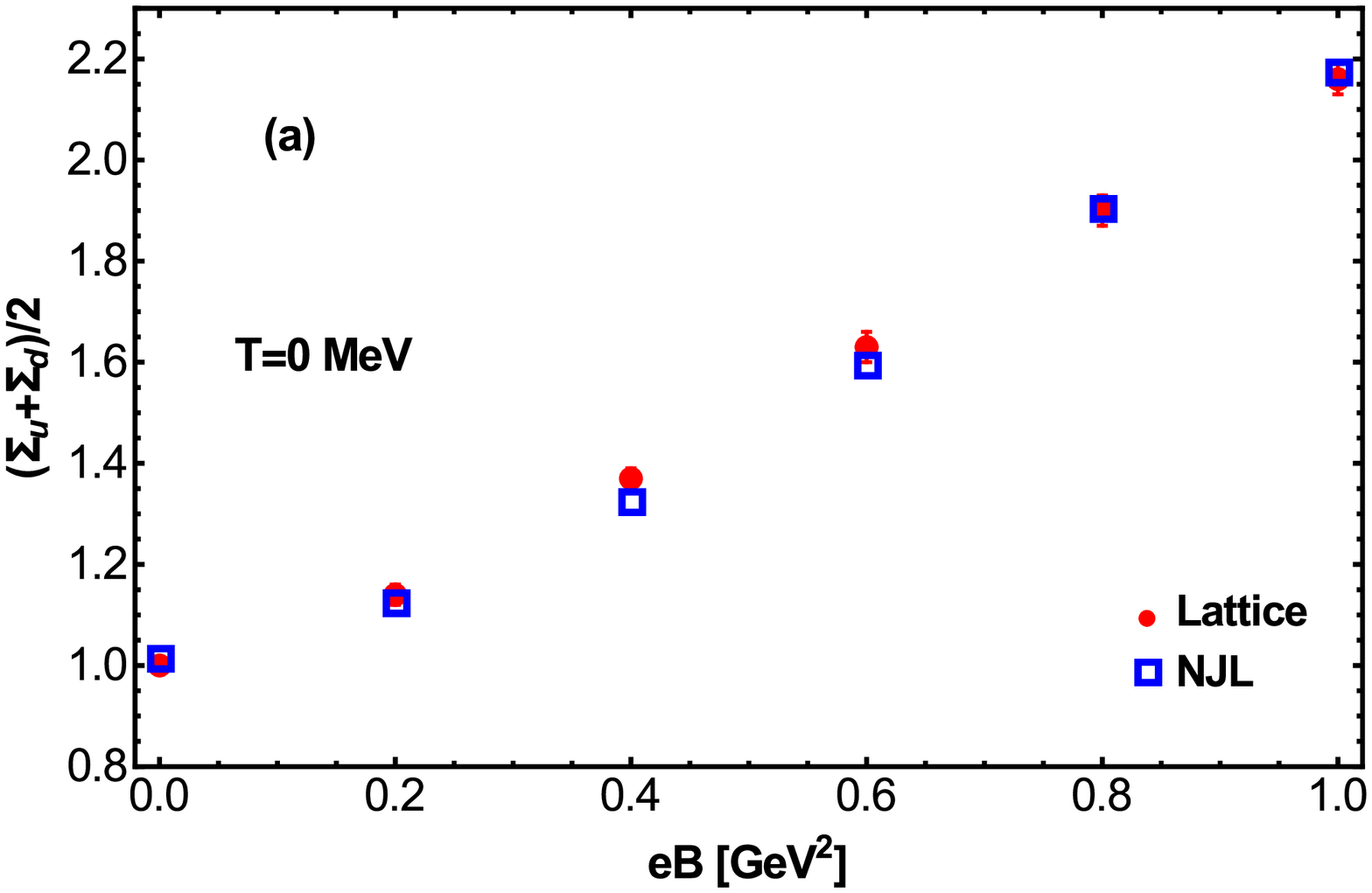}
\includegraphics[scale=0.42]{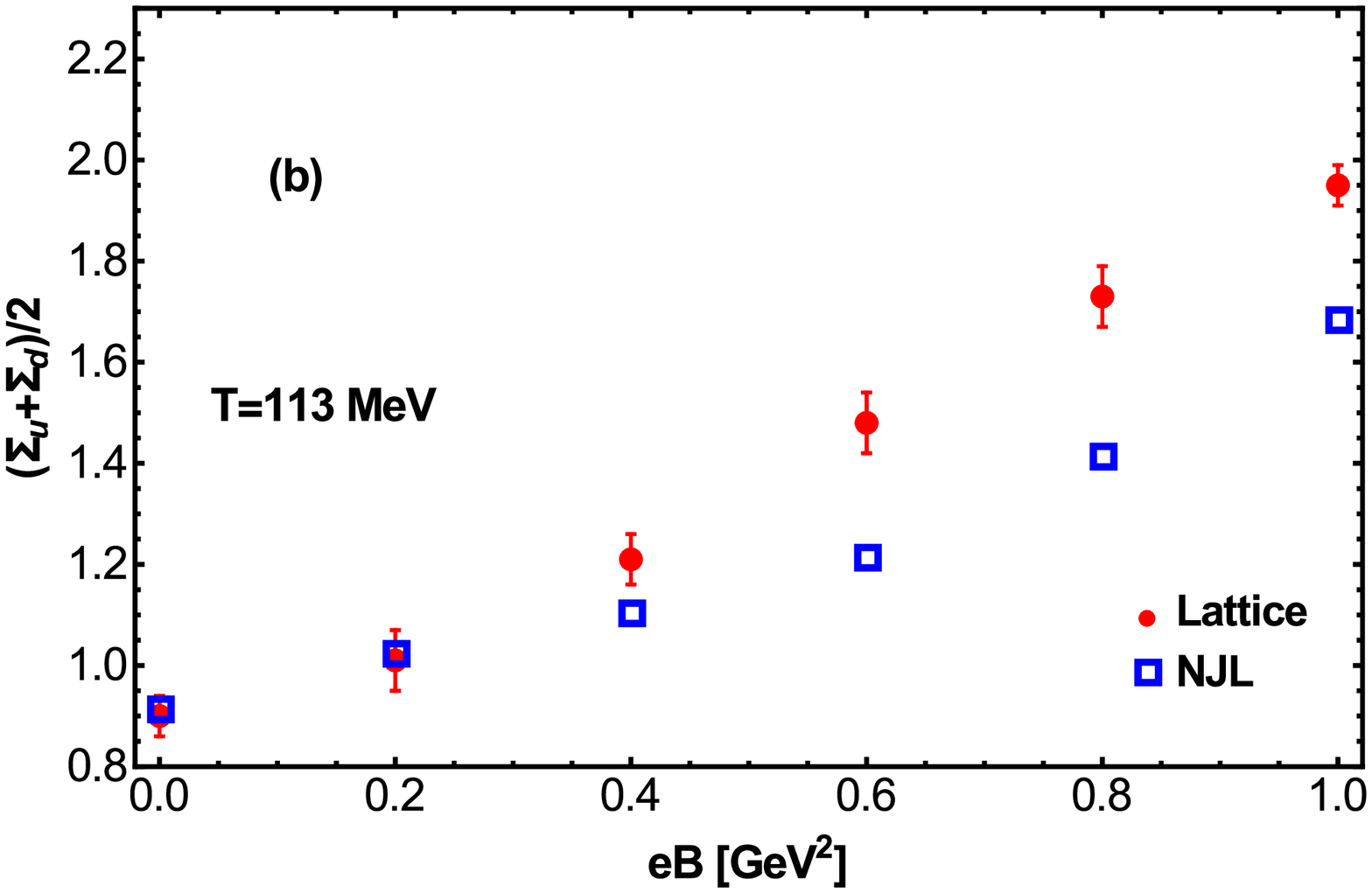}
\end{center}
\begin{center}
\includegraphics[scale=0.42]{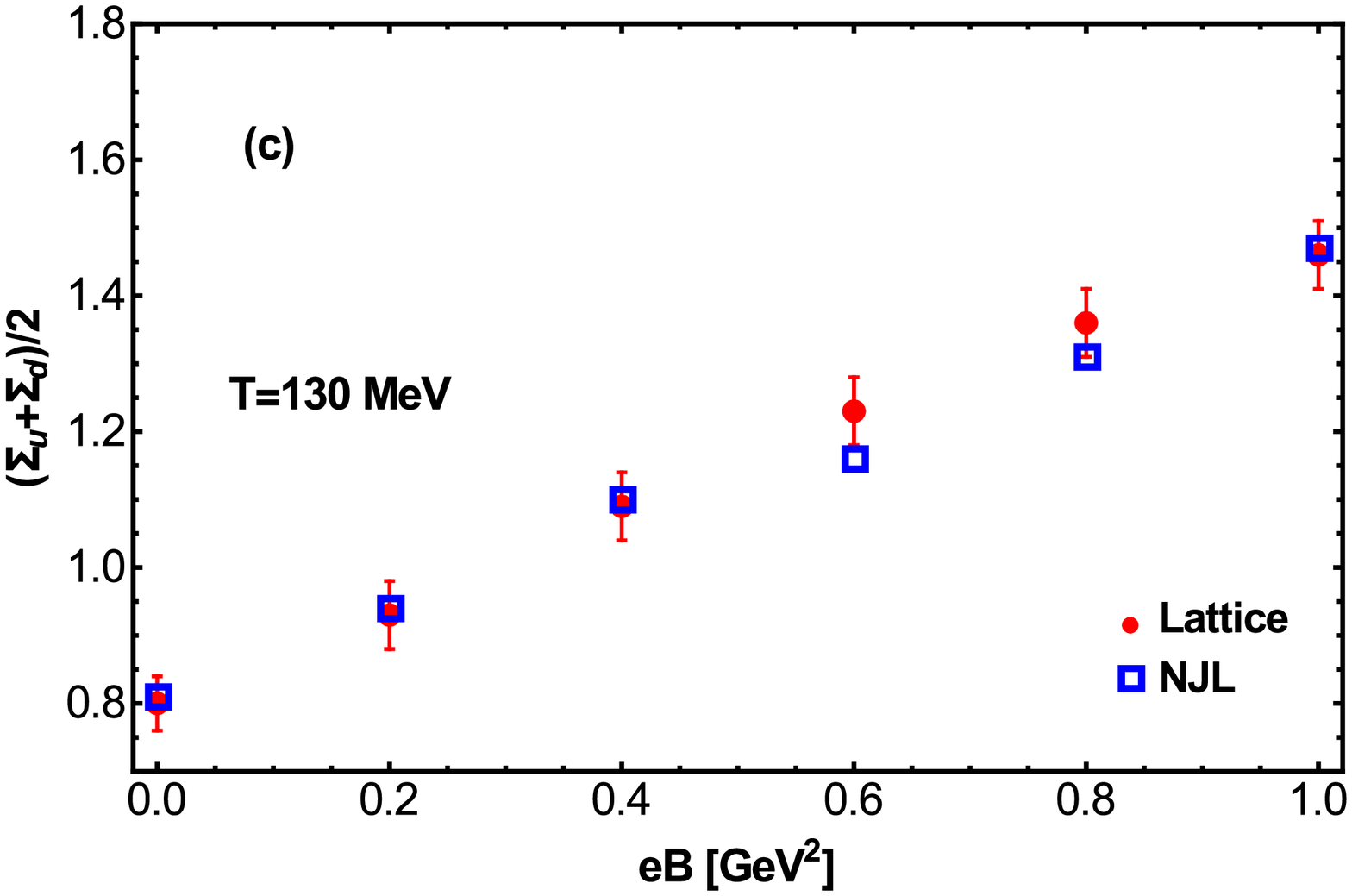}
\includegraphics[scale=0.42]{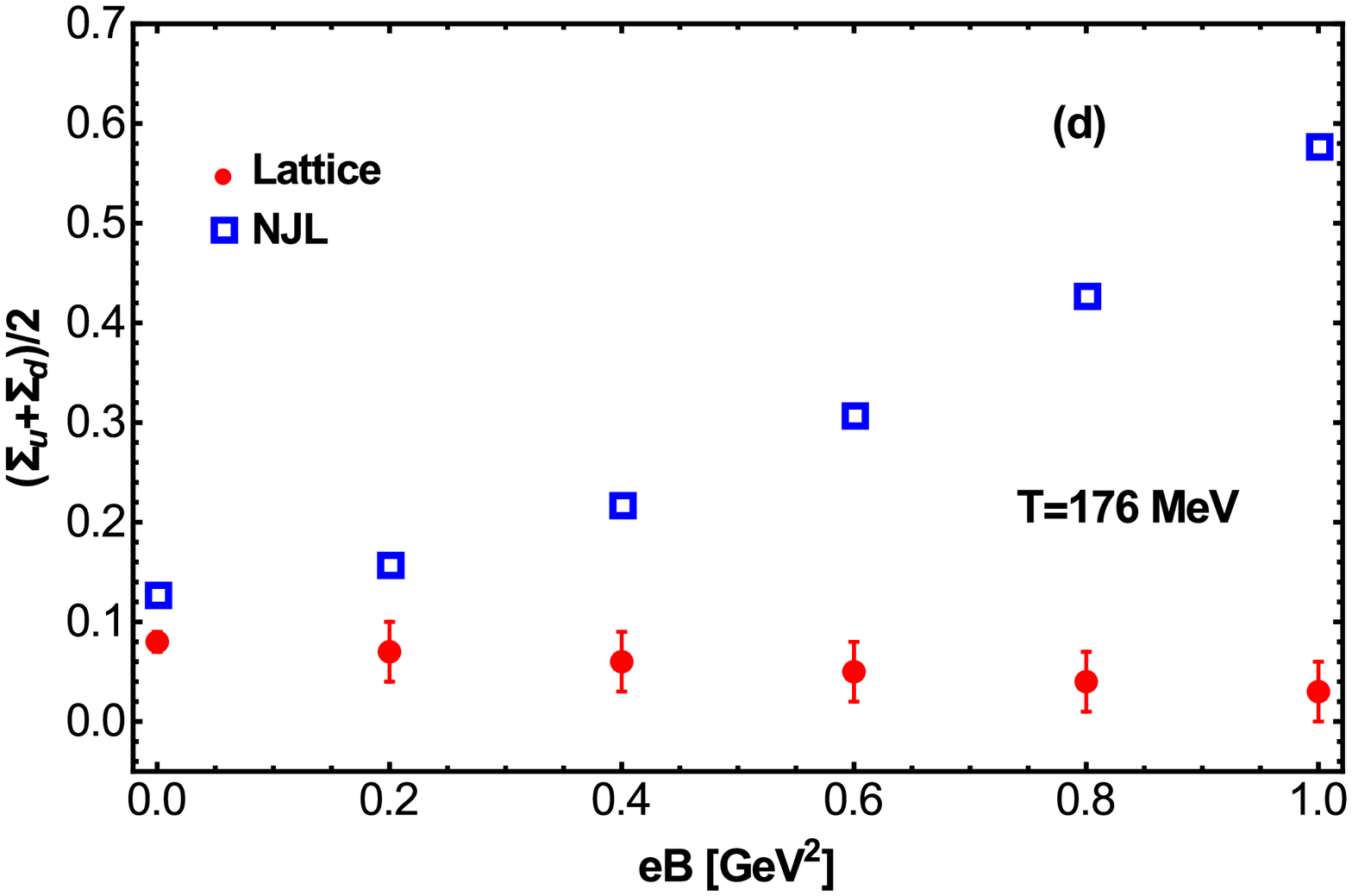}
\end{center}
\caption{Comparison between the lattice QCD results from Ref.~\cite{bali2} for the average quark condensate and the model calculation, as a function of the magnetic field. Curves $(a)$, $(b)$, $(c)$ and $(d)$ correspond to  $T=0,\ 113,\  130$ and $176\, {\mbox{MeV}}$, respectively. The model describes better the lattice results for lower temperatures.}
\label{fig1}
\end{figure*}
\end{widetext}

We use as the condition to find the critical temperature the vanishing of the derivative of the thermal piece of the quark condensate with respect to the temperature, namely
\begin{eqnarray}
   \frac{d}{dT}\langle\bar{\psi}\psi\rangle_{0,T}=0,
\label{derivative}
\end{eqnarray} 
where
\bea
   \!\!\!\!\!\!\!\!\!\!\langle\bar{\psi}\psi\rangle_{0,T}=\frac{2N_cM}{\pi^2}
   \int_0^\infty \!\!\!\!dp\frac{p^2}{\sqrt{p^2+M^2}}\frac{1}{e^{\sqrt{p^2+M^2}/T}+1}.
\label{thermalcond}
\eea
\\

A straightforward calculation in the approximation where $T_c^{NJL}$ and $M$ are of the same order gives
\begin{eqnarray}
  T_c^{NJL}\simeq 2.38 M.
\label{straight}
\end{eqnarray} 

To have a better estimate of $T_c^{NJL}$, one needs to evaluate the above equation at an appropriate value of $M$. We observe that at the critical temperature, the dynamically generated mass drops from its vacuum value to about half of it, namely to $M_0/2$. Using this as the working criterium and the reported LQCD value for the critical temperature $T_c=0.158$ GeV, we obtain the corresponding values for the model critical temperature, which we also show in Table~1.

\section{thermo-magnetic behavior of the dynamically generated masses and couplings}\label{IV}

\begin{figure}[b]
\begin{center}
\includegraphics[scale=0.42]{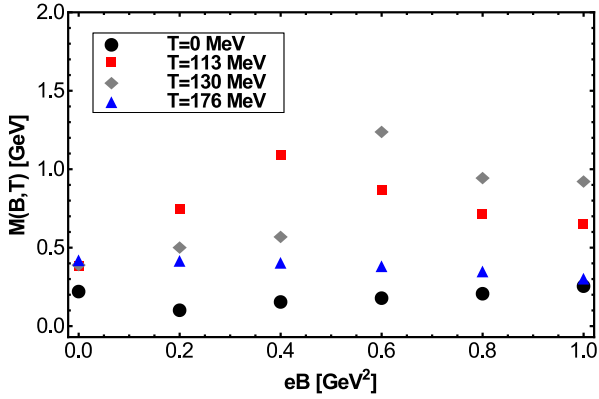}
\end{center}
\caption{Thermo-magnetic average quark mass $M(B,T)$ as a function of the field strength $eB$ for the temperatures $T=0\, {\mbox{MeV}},\ 113\, {\mbox{MeV}},\ 130\, {\mbox{MeV}}$ and $176\, {\mbox{MeV}}$ computed using the first set of values in Table 1.}
\label{fig2}
\end{figure}
To establish how well the solutions for $M(B,T)$ describe the condensates as functions of $eB$ and $T$ one considers Fig.~\ref{fig1}. The figure shows the LQCD average quark condensate $(\Sigma_u + \Sigma_d)/2$ compared to the equivalent quantity $1 + \langle\bar{\psi}\psi\rangle_{B,T}/\langle\bar{\psi}\psi\rangle_0$ computed within the model, using Eq.~(\ref{condpropthermmag}) for $\langle\bar{\psi}\psi\rangle_{B,T}$ and one of the values in Table 1 for $\langle\bar{\psi}\psi\rangle_0$. 

\begin{figure}[b]
\begin{center}
\includegraphics[scale=0.42]{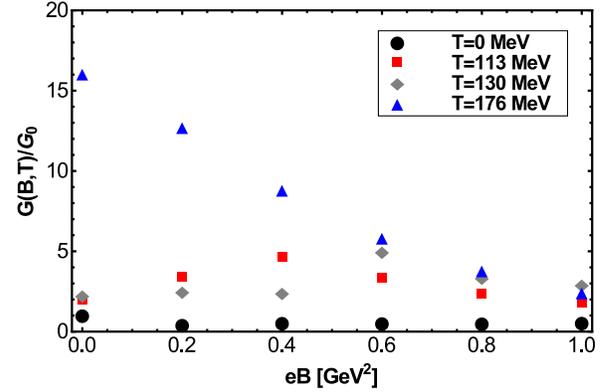}
\end{center}
\caption{Thermo-magnetic average coupling $G(B,T)$ as a function of the field strength $eB$ for the temperatures $T=0\, {\mbox{MeV}},\ 113\, {\mbox{MeV}},\ 130\, {\mbox{MeV}}$ and $176\, {\mbox{MeV}}$ computed using the first set of values in Table 1.}
\label{fig3}
\end{figure}

Equation~(\ref{condpropthermmag}) is a transcendental equation with none, one or multiple solutions for $M$, depending on the values of $T$ and $B$. The procedure we follow to find the reported value of $M$ is to average the multiple solutions in the case that there is more than one or to define as the solution the value of $M$ that provides the closest distance between the lattice value and the model. From Fig.~\ref{fig1} we notice that our description of the LQCD results is better for low temperatures. Figure~\ref{fig1} has been prepared using the first set of values in Table 1.

The behavior of the thermo-magnetic average mass $M(B,T)$ and coupling $G(B,T)$ as functions of the field strength are depicted in Figs.~\ref{fig2} and~\ref{fig3}, respectively. Notice that for $T=0$, the mass increases monotonically with the magnetic field. However, there is a turn-over behavior for intermediate values of $T$ where, as functions of $eB$ the masses start off increasing to then decrease as the field strength increases. For the largest temperature, which is above the transition temperature, the mass becomes a monotonically decreasing function of the field strength. A similar behavior is observed for the coupling. Although $G(B,0)$ is practically constant, for the temperature above the transition temperature the coupling becomes a monotonically decreasing function of the field strength. For intermediate temperatures there is also a a turn-over behavior where as functions of $eB$ the couplings start off increasing and then decrease as the field strength increases.

\begin{figure}[b]
\begin{center}
\includegraphics[scale=0.42]{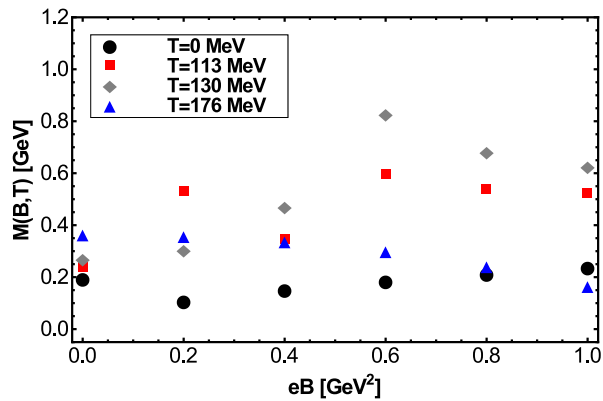}
\end{center}
\caption{Thermo-magnetic average quark mass $M(B,T)$ as a function of the field strength $eB$ for the temperatures $T=0\, {\mbox{MeV}},\ 113\, {\mbox{MeV}},\ 130\, {\mbox{MeV}}$ and $176\, {\mbox{MeV}}$ computed using the second set of values in Table 1.}
\label{fig4}
\end{figure}

\begin{figure}[b]
\begin{center}
\includegraphics[scale=0.42]{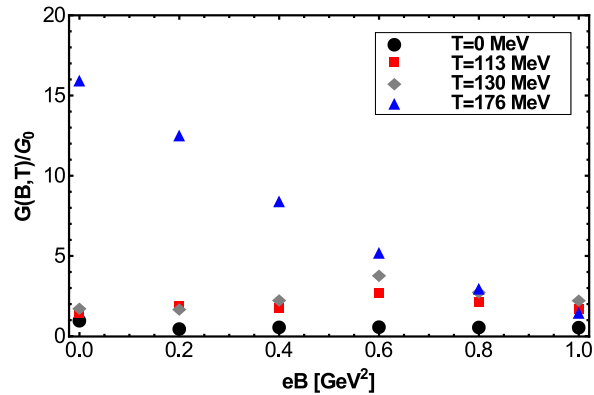}
\end{center}
\caption{Thermo-magnetic average coupling $G(B,T)$ as a function of the field strength $eB$ for the temperatures $T=0\, {\mbox{MeV}},\ 113\, {\mbox{MeV}},\ 130\, {\mbox{MeV}}$ and $176\, {\mbox{MeV}}$ computed using the second set of values in Table 1.}
\label{fig5}
\end{figure}

To test the sensitivity of the results to a change in the vacuum parameters, Figs.~\ref{fig4} and~\ref{fig5} show the behavior of the $M(B,T)$ and $G(B,T)$ as functions of the field strength for different temperatures, when using the second set of values in Table 1 for the calculation. Notice that the results are qualitative and quantitatively similar to the ones obtained from the first set of values in Table 1.

In order to study one of the consequences of the behavior of the mass and coupling, we proceed to compute the thermo-magnetic contribution to the pressure. Notice that the magnetic field induces a difference between the pressure in the directions parallel and perpendicular to the field; a magnetization in the former direction is absent, while in the latter it contributes. We call the first kind of pressure {\it longitudinal}, that is, directed along the $\hat{z}$ axis, whereas we call the second kind of pressure {\it transverse}, that is, directed along the $\hat{x}$, $\hat{y}$ directions. We consider only the renormalized contribution to the pressure in the so called \lq\lq$\Phi$-scheme\rq\rq~\cite{bali3}. In the mean field approximation, the longitudinal contribution to the pressure can be written as 
\bea
P_z=-V^{\mathrm{\mbox{\small{eff}}}},
\label{press}
\eea
where $V^{\mathrm{\mbox{\small{eff}}}}$, is the effective potential. Therefore, using Eqs.~(\ref{gap1}) and~(\ref{qbarq}), $P_z$ can be written as
\begin{eqnarray}
P_z&=&-\frac{(M(B,T)-m)^2}{4G}\nonumber\\
&-& \frac{i}{2}\sum_f \mbox{Tr} \int \frac{d^4p}{(2\pi)^4} \ln
(iS_f^{-1}),
\label{pressure}
\end{eqnarray}
whereas the magnetization $\vec{\mathcal{M}}$ is given by
\bea
   \vec{\mathcal{M}}=-\frac{\partial V^{\mathrm{\mbox{\small{eff}}}}}{\partial (eB)}\hat{z},
\label{magnetization}
\eea 
from where the transverse pressure can be computed as~\cite{bali3}
\bea
  P_{x,y}=P_z + e\vec{B}\cdot\vec{\mathcal{M}}.
\label{longpress}
\eea
To compute $\vec{\mathcal{M}}$, we observe that $M(B,T)$ has a mild dependence on $eB$. Therefore, we only consider the terms coming from the explicit dependence of $eB$ of the effective potential and of $G(B,T)$, which are by far the dominant contributions.

\begin{figure}[t!]
\begin{center}
\includegraphics[scale=0.41]{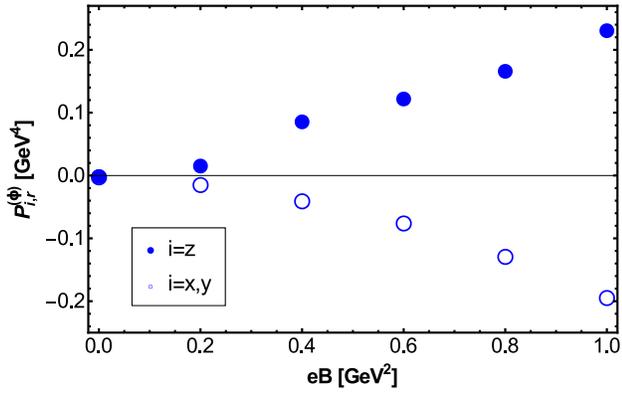}
\end{center}
\caption{Longitudinal and transverse pressures as functions of the field strength $eB$ for $T=113\, {\mbox{MeV}},\ $ computed using the first set of vacuum values in Table 1. The longitudinal (transverse) pressure is a monotonically increasing (decreasing) function that for this temperature starts off from zero and grows (decreases) towards positive (negative) values as the field strength increases.}
\label{fig6}
\end{figure}

Notice that for the computation of the pressure we use $M(B,T)$ and $G(B,T)$, namely, the average mass and coupling, respectively. Therefore, the pressure and magnetization are correspondingly also computed as an average over the light flavors.

\begin{figure}[t!]
\begin{center}
\includegraphics[scale=0.42]{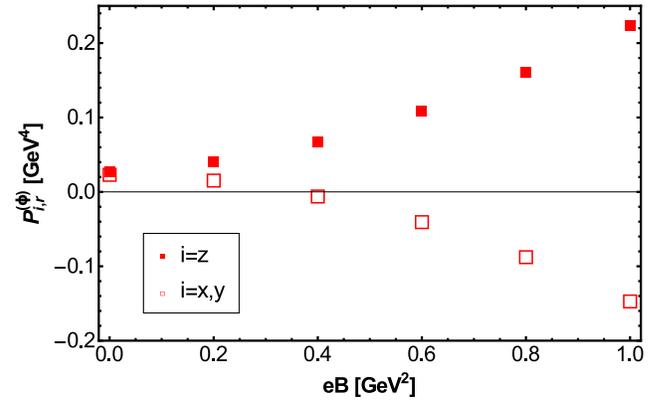}
\end{center}
\caption{Longitudinal and transverse pressures as functions of the field strength $eB$ for $T=176\, {\mbox{MeV}},\ $ computed using the first set of vacuum values in Table 1. The longitudinal (transverse) pressure is a monotonically increasing (decreasing) function that for this temperature starts off from positive values and grows (decreases) towards positive (negative) values as the field strength increases.}
\label{fig7}
\end{figure}

Figures~\ref{fig6} and~\ref{fig7} show the longitudinal and transverse pressures as functions of the field strength, for $T=113$ MeV and $T=176$ MeV, respectively, computed using the first set of vacuum parameters in Table 1. Notice that for $T=113$ MeV, that is, for a temperature below $T_c$, these pressures start off from zero and have opposite behaviors; the longitudinal pressure is a monotonically increasing function towards positive values whereas the transverse pressure is a monotonically decreasing function towards negative values. For the case of $T=176$ MeV, that is for a temperature above $T_c$, both pressures start off from positive values. However, there is an interval of field strengths where the transverse pressure is positive to then change sign and become negative. This results are in agreement with the findings of Ref.~\cite{bali3}. 

\begin{figure}[t!]
\begin{center}
\includegraphics[scale=0.41]{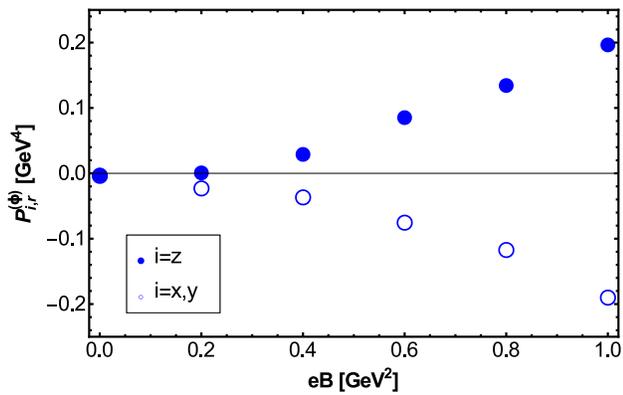}
\end{center}
\caption{Longitudinal and transverse pressures as functions of the field strength $eB$ for $T=113\, {\mbox{MeV}},\ $ computed using the second set of vacuum values in Table 1. The results are equivalent to the ones obtained using the first set of vacuum values in Table 1.}
\label{fig8}
\end{figure}

\begin{figure}[t]
\begin{center}
\includegraphics[scale=0.42]{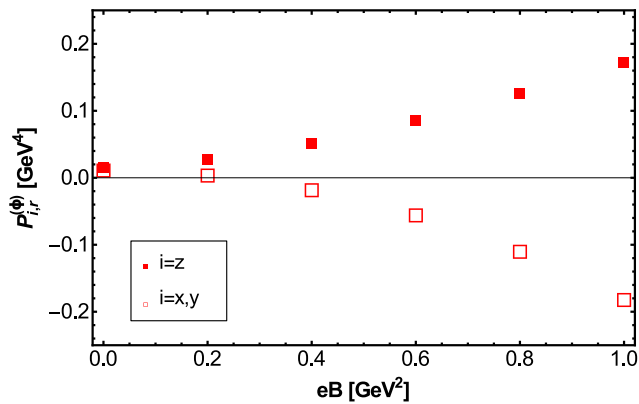}
\end{center}
\caption{Longitudinal and transverse pressures as functions of the field strength $eB$ for $T=176\, {\mbox{MeV}},\ $ computed using the second set of vacuum values in Table 1. The results are equivalent to the ones obtained using the first set of vacuum values in Table 1.}
\label{fig9}
\end{figure}

To test the sensitivity of the results to the vacuum parameters, Figs.~\ref{fig8} and~\ref{fig9} show the same pressures computed using the second set of parameters in Table 1. The results are equivalent.

It is important to notice that the calculation describing the longitudinal pressure in Fig.~\ref{fig6} -- Fig.~\ref{fig9} agree with the LQCD findings~\cite{bali3}, {\it provided} that
\bea
   e\vec{B}\cdot\vec{\mathcal{M}}=-eB{\mathcal{M}},
\label{provided}
\eea
where ${\mathcal{M}}$ represents the magnitude of the magnetization vector. This means, that the magnetization is overall opposite to the magnetic field which in turn means that the system is well described in the model as possessing diamagnetic properties, both, below and above the critical temperature.

\begin{figure}[t!]
\begin{center}
\includegraphics[scale=0.41]{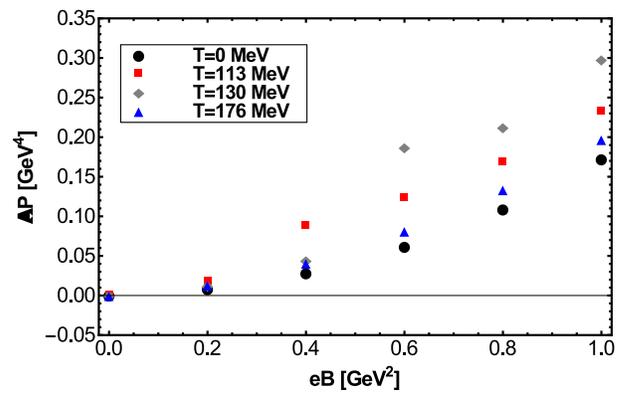}
\end{center}
\caption{$\Delta P\equiv P_z(B,T) - P_z(0,T)$, computed for the first set of vacuum values in Table 1 for $T=0,\ 113,\ 130$ and $176$ MeV.}
\label{fig10}
\end{figure}

\begin{figure}[t!]
\begin{center}
\includegraphics[scale=0.42]{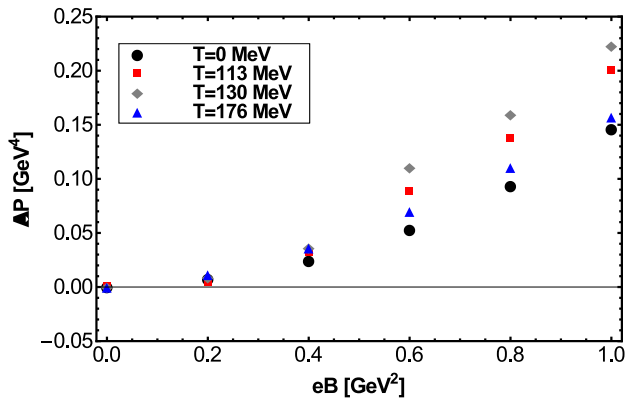}
\end{center}
\caption{$\Delta P\equiv P_z(B,T) - P_z(0,T)$, computed for the second set of vacuum values in Table 1 for $T=0,\ 113,\ 130$ and $176$ MeV.}
\label{fig11}
\end{figure}

Finally, in order to study the pressure behavior referred to purely thermal effects, Figs.~\ref{fig10} and~\ref{fig11} show $\Delta P\equiv P_z(B,T) - P_z(0,T)$, computed for the first and second set of vacuum values in Table 1, respectively. In all cases $\Delta P$ is well described by a monotonically and positive definite increasing function of $eB$. This behavior is also in agreement with LQCD calculations~\cite{bali3}. We observe that the rate of change shows a turn over behavior as the temperature increases. For $T=0$ the rate of increase is small, becoming faster for intermediate temperatures to then decrease for the highest temperature.
\\

\section{Summary and conclusions}\label{concl}

In this paper we studied the thermo-magnetic behavior of the coupling constant and mass in the NJL model. We used the gap equation and the quark condensate expressions obtained from the model together with LQCD results for the light-quark condensates in the presence of a magnetic field~\cite{bali2}. Although we obtained the behavior of the couplings $G(B,T)$ and $M(B,T)$ as a function of the temperature and the magnetic field strength, we did not attempt a detailed magnetic field dependence  study of a list of observables, neither to look for a parametrization of the coupling as a function of the temperature and/or the field strength. A study of this sort has been recently carried out in Ref.~\cite{Krein}.

Our results show that for temperatures above the transition temperature, the couplings are monotonically decreasing functions of the field strength. This means that at these temperatures the melting of the quark condensates is accompanied by a corresponding decrease in the strength of the interaction that binds these quarks.

For temperatures close to, but below the transition temperature, we find a turn over behavior of the couplings. As the field strength starts increasing, the couplings increase. However, for intermediate values of the field strength the couplings decrease. This signals that as the temperature decreases below, but close to the transition temperature, the strength of the coupling increases. This increase is accompanied by a corresponding increase in the value of the condensate, as shown by LQCD calculations. Nevertheless, this increase is not sustained, since for stronger fields the couplings decrease, as do the LQCD computed condensates. 

The results for $G(B,T$) strengthen the picture advocated in Refs.~\cite{coupling} where the behavior of the condensate as a function of the magnetic field is directly linked to the properties of the strong coupling constant at high and low temperatures.

We also computed the thermo-magnetic contribution to the longitudinal and transverse pressures. We found that below $T_c$, the transverse pressure as a function of the magnetic field, decreases towards negative values starting off from zero. However, for temperatures above the transition temperature, although the transverse pressure still decreases as a function of the field strength, it starts off from positive values. This turnover behavior of the transverse pressure means that above $T_c$ particles are pulled closer together, at least for small values of the magnetic field. The fact that at the same time the coupling decreases can be viewed as signaling that the strength of the bound of the condensate is smaller, due to asymptotic freedom and this can be responsible for the decrease of the condensate as the magnetic field strength is turned on. Last but not least, we found that in order for the computation of the longitudinal pressure to agree with LQCD calculations, the system should be described as a diamagnet at finite temperature, that is, with its magnetization opposite to the magnetic field direction, both above and below the critical temperature.

Overall, the results suggest that IMC, as described by the thermo-magnetic behavior of the quark condensate, can be linked to the properties of the coupling constant as a function of the magnetic field in a wide range of temperatures.

\section*{Acknowledgments}

A.A. and L.A.H. acknowledge valuable discussions with N. Scoccola. This work has been supported in part  by  UNAM-DGAPA-PAPIIT grant number IN101515, by Consejo Nacional de Ciencia y Tecnolog\1a grant number 256494, by CIC-UMSNH (M\'exico) grant number 4.22,  National Research Foundation (South Africa), the Harry Oppenheimer Memorial Trust OMT Ref. 20242/02 and  by FONDECYT (Chile) grant numbers 1130056, 1120770, 1150471 and 1150847.  CV acknowledges support from the group {\em F\1sica de Altas Energ\1as} at UBB. L.A.H. acknowledges the University of Cape Town and the National Research Foundation (South Africa) for funding assistance. M.L acknowledges support from Proyecto Basal (Chile) FB 0821.

\end{document}